\documentclass[aps,prl,twocolumn,groupedaddress,showpacs,superscriptaddress]{revtex4-2}

\usepackage[normalem]{ulem}

\usepackage[latin1]{inputenc}
\usepackage{graphicx}
\usepackage{float}
\usepackage{latexsym}
\usepackage{graphicx}
\usepackage{amssymb}
\usepackage{amsmath}
\usepackage{amsfonts}
\usepackage[colorlinks=true,citecolor=blue,linkcolor=blue,urlcolor=magenta]{hyperref}

\usepackage{graphicx}
\usepackage{dcolumn}
\usepackage{bm}

\usepackage{dcolumn}
\usepackage{bm}
\usepackage{amssymb, graphicx, amsmath}
\usepackage{mathdots}
\usepackage{mathrsfs}
\usepackage[usenames,dvipsnames]{color}
\usepackage{subfigure}
\usepackage{float}
\usepackage{verbatim}
\usepackage{hyperref}
\usepackage{graphicx}
\usepackage{tikz}
\usetikzlibrary{quantikz}
\usepackage{soul}
\usepackage{comment}

\newcommand{\fig}[1]{Fig.\ \ref{#1}}

\begin{document}

\title{Characterization and Optimization of Tunable Couplers via Adiabatic Control in Superconducting Circuits}

\author{Xuan Zhang}
\altaffiliation{These authors contributed equally to this work.}
\affiliation{Department of Physics and Guangdong Basic Research Center of Excellence for Quantum Science, Southern University of Science and Technology (SUSTech), Shenzhen 518055, China}

\author{Xu Zhang}
\altaffiliation{These authors contributed equally to this work.}
\affiliation{Department of Physics and Guangdong Basic Research Center of Excellence for Quantum Science, Southern University of Science and Technology (SUSTech), Shenzhen 518055, China}

\author{Changling Chen}
\affiliation{Shenzhen Institute for Quantum Science and Engineering , Southern University of Science and Technology, Shenzhen 518055, China}

\author{Kai Tang}
\affiliation{Department of Physics and Guangdong Basic Research Center of Excellence for Quantum Science, Southern University of Science and Technology (SUSTech), Shenzhen 518055, China}

\author{Kangyuan Yi}
\affiliation{Department of Physics and Guangdong Basic Research Center of Excellence for Quantum Science, Southern University of Science and Technology (SUSTech), Shenzhen 518055, China}

\author{Kai Luo}
\affiliation{Department of Physics and Guangdong Basic Research Center of Excellence for Quantum Science, Southern University of Science and Technology (SUSTech), Shenzhen 518055, China}

\author{Zheshu Xie}
\affiliation{Department of Physics and Guangdong Basic Research Center of Excellence for Quantum Science, Southern University of Science and Technology (SUSTech), Shenzhen 518055, China}

\author{Yuanzhen Chen}
\email{chenyz@sustech.edu.cn}

\affiliation{Department of Physics and Guangdong Basic Research Center of Excellence for Quantum Science, Southern University of Science and Technology (SUSTech), Shenzhen 518055, China}

\author{Tongxing Yan}
\email{yantx@sustech.edu.cn}

\affiliation{Shenzhen Institute for Quantum Science and Engineering , Southern University of Science and Technology, Shenzhen 518055, China}

\affiliation{International Quantum Academy, Shenzhen 518048, China}
\date{\today}

\begin{abstract}
In the pursuit of scalable superconducting quantum computing, tunable couplers have emerged as a pivotal component, offering the flexibility required for complex quantum operations of high performance. In most current architectures of superconducting quantum chips, such couplers are not equipped with dedicated readout circuits to reduce complexity in both design and operation. However, this strategy poses challenges in precise characterization, calibration, and control of the couplers. In this work, we develop a hardware-efficient and robust technique based on adiabatic control to address the above issue. The critical ingredient of this technique is adiabatic swap (aSWAP) operation between a tunable coupler and nearby qubits. Using this technique, we have characterized and calibrated tunable couplers in our chips and achieved straightforward and precise control over these couplers. For example, we have demonstrated the calibration and correction of the flux distortion of couplers. In addition, we have also expanded this technique to tune the dispersive shift between a frequency-fixed qubit and its readout resonator over a wide range. 

\end{abstract}

\maketitle

Superconducting circuits based on circuit quantum electrodynamics (cQED) are a leading platform for quantum computing \cite{Blais_MRP,Acharya2024}. Among the various architectures belonging to this platform \cite{Kim2023,Arute_nature}, those containing tunable couplers have been extensively explored, achieving impressive progress in recent years \cite{Arute_nature,Yulinprl, Stehlik_PRL,Sungprx}. Tunable couplers offer several key advantages for scaling up superconducting circuits\cite{Yan2018prappld}. Specifically, flexible, wide-range tuning of the coupling between qubits can be conveniently realized by modulating the coupler frequency, which is crucial for high-fidelity gate operations \cite{Arute_nature, Sungprx,Foxen_prl}. In addition, the issue of frequency crowding and crossing among qubits often encountered in other architectures can be largely mitigated since only couplers need to be scanned over a broad frequency range during gate operations \cite{Zhang2022,Xu2020prl}. 

However, such advantages do come at the cost of hardware overhead, as tunable couplers require additional control \cite{Sungprx}. A common compromise is to minimize the number of supporting components for these couplers, such as dedicated readout circuits \cite{Arute_nature, Yulinprl}. Nevertheless, precisely characterizing the couplers becomes challenging without such components. One way to circumvent this issue is to design readout resonators shared by the couplers and their neighboring qubits \cite{Collodo_prl}, which not only complicates circuit design but also introduces additional decoherence channels to the qubits. Another approach is to indirectly measure the states of the couplers by utilizing the strong dispersive coupling between them and nearby qubits \cite{Xu2020prl,Zhangapl}. However, in this case, the duration of the readout pulses is typically rather long, making decoherence a potential bottleneck. A hardware-efficient yet reliable characterization of tunable couplers is thus highly desirable for scalable superconducting quantum computation.

In this article, we propose and demonstrate a hardware-efficient and robust scheme to characterize and calibrate such tunable couplers based on adiabatic control, which we call adiabatic SWAP (aSWAP). The scheme starts with adiabatically swapping the states of a coupler and a neighboring qubit equipped with a dedicated readout resonator, then, the state of the coupler is inferred by measuring the qubit. Using this straightforward method, the basic parameters of the coupler, including its frequency and decoherence times, can be accurately characterized at arbitrary bias points. Furthermore, this method is also employed to calibrate flux distortion of couplers with high precision, which is critical for enhancing the performance of two-qubit gates. Finally, the effective dispersive shift between a frequency-fixed qubit and its readout resonator can be modulated in a wide range by aSWAP. This provides an alternative approach to facilitating readout optimization and supporting applications that require a tunable dispersive shift.

\begin{figure}
    \centering
    \includegraphics[width=\linewidth]{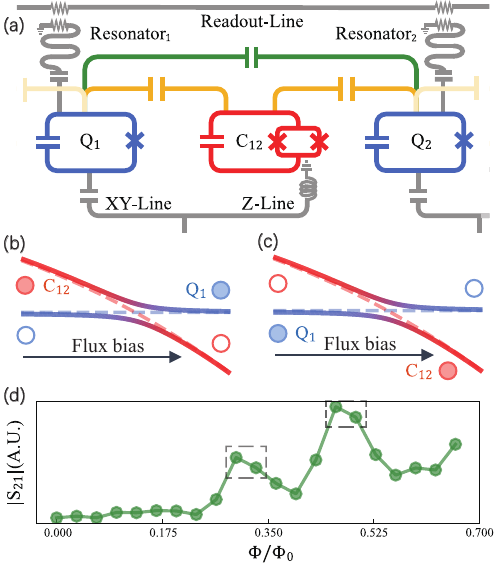}
    \caption{(a) Partial schematic of a superconducting quantum circuit with tunable coupling. $Q_1$ and $Q_2$ are two qubits with fixed frequencies and $C_{12}$ is a tunable coupler (in this work, $f_{Q_1}$ = 4.636 GHz, $f_{Q_2}$ = 4.127 GHz). All three components are capacitively coupled to each other. Each qubit has its own readout resonator. (b)\&(c) Illustration of two types of aSWAP operations used in this work. The dashed lines indicate bare energy levels of the coupler (red) and the qubit (blue). Near the two crossing points of the dashed lines, strong coupling between the coupler and the qubit hybridizes their states and lifts the degeneracy, resulting in the spectra of dressed states (solid lines). The color gradient in the solid lines indicates the relative weight of the bare states in the dressed states. The filled and hollow circles represent first-excited and ground states, respectively. (d) Transmission coefficient of the readout pulse as the flux bias of the coupler is varied. The two peaks correspond to two anticrossing points where the coupler comes close to resonance with the two qubits.}
    \label{circuit}
\end{figure}

We first discuss the principle of our scheme. Recently, a similar idea was also exploited for resetting qubits and leakage reduction \cite{chen2024leakage}. Figure\ \ref{circuit}(a) schematically shows part of a typical cQED circuit composed of two frequency-fixed qubits and a tunable coupler. The coupler is a tunable transmon whose frequency can be adjusted via magnetic flux \cite{luokai2023}. By changing the frequency of the coupler, the effective coupling between the two qubits can be varied as desired. Naturally, the accuracy of such tunability relies on a precise characterization of the coupler. 

When the coupler is far-detuned from both qubits, the hybridization between the three components can be neglected. As the frequency of the coupler is tuned close to that of a qubit, their states hybridize, and a lifted degeneracy occurs near resonance, as illustrated in \fig{circuit}(b)\&(c). According to the adiabatic theorem, when the coupler is tuned slowly enough, any eigenstate (initially having a dominant weight on either a qubit or the coupler) remains in the instantaneous eigenstate throughout the flux sweep. Consequently, when passing through a lifted degeneracy,  the states of the coupler and the qubit swap. Building on this principle, we can implement two different aSWAP operations: one for readout (\fig{circuit}(b)) and the other for state preparation (\fig{circuit}(c)) of the coupler. 

From the above discussion, it is clear that successfully implementing the aSWAP technique requires identifying the anticrossing points of the energy levels of the coupler and its nearby qubits. The approximate positions of the anticrossing points can be easily located by several methods and one example is given in \fig{circuit}(d). In this measurement, a readout signal at a frequency optimized for the resonator of $Q_1$ is sent through the readout line while the flux bias of the coupler is varied. At zero bias, the coupler's frequency is much higher than those of the two qubits. Increasing the bias lowers the coupler's frequency. When the coupler approaches resonance with either qubit, an anticrossing of their energy levels occurs, causing a change in the amplitude of the transmission coefficient of the readout pulse. Notice that due to unavoidable stray couplings between $Q_2$ and $Q_1$ and their resonators in the planar layout of these components, a signal also appears as the coupler's frequency comes close to that of $Q_2$, even though there is no direct capacitive coupling between them by design. 

\begin{figure}[t]
    \centering
    \includegraphics[width=\linewidth]{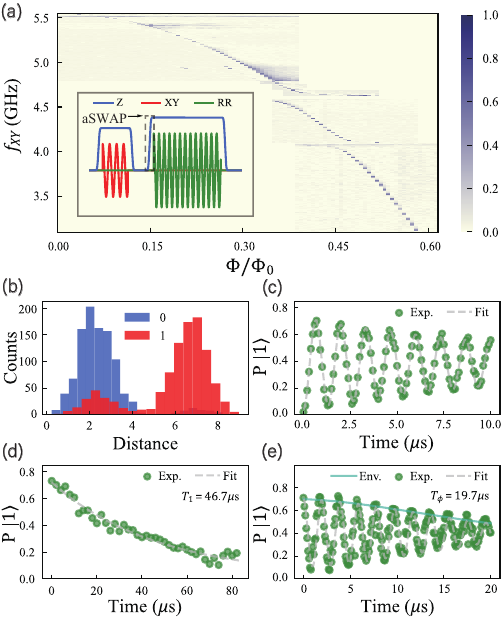}
    \caption{Characterization of the coupler $C_{12}$. (a) Spectroscopy of the coupler's frequency as a function of its flux bias. The inset shows the measurement sequence of pulses applied to the $Z$-line, $XY$-line, and the readout resonator, respectively (see \fig{circuit}(a)). The $Z$ pulse sets the magnetic flux $\Phi$ and thus the frequency of the coupler. The $XY$ pulse is felt by both the coupler and the qubit $Q_1$ (see \fig{circuit}(a)). To achieve a complete aSWAP operation, one needs to make sure that the frequency of the coupler be tuned in a range wide enough covering an anticrossing point. With the knowledge of approximate positions of the anticrossing points obtained in \fig{circuit}(d), the above requirement can always be fulfilled. (b) Histogram of measuring the state of $C_{12}$ using the readout resonator of $Q_1$ via an aSWAP operation. (c) Rabi oscillation on $C_{12}$ at its highest frequency. (d) Relaxation process of $C_{12}$ at its highest frequency. (e) Ramsey interference on $C_{12}$ at its highest frequency.}
    \label{spectrum}
\end{figure}

After determining the approximate flux bias values of the two anticrossing points, we can now perform a detailed measurement to map out the full energy spectrum of the $Q_1-C_{12}-Q_2$ system. The measurement sequence and results are shown in \fig{spectrum}(a). A $Z$-line signal first brings the coupler from its idling point ($\Phi=0$, highest frequency) down to a specific biasing point, $\Phi$. Then, a driving pulse with a frequency of $f_{XY}$ is sent through the $XY$-line. Next, an aSWAP operation between the coupler and its nearby qubit is performed before the readout measurement. The coupler is excited when $f_{XY}$ is resonant with the coupler and the excitation will be swapped to one of the qubits (depending on which adiabatic branch the instantaneous eigenstate is on) with the subsequent aSWAP operation, which yields a nonzero population in the readout result. We repeat the above sequence at different biases to obtain the full spectrum. Within the vicinity of the two anticrossing points, a lifted degeneracy is observed. The energy splitting at these points can be used to determine the coupling strength between the coupler and the qubits.

Here we want to emphasize a convenient feature of the aSWAP technique for various applications discussed later. While the full spectrum in \fig{spectrum}(a) reveals the precise position of the anticrossing points, such information is not necessary for all subsequent measurements. If we were to realize a high-fidelity gate operation of SWAP type between a coupler and a qubit, two conditions must indeed be fulfilled. One is knowing the position of the relevant anticrossing point, and the other one is the adiabaticity during aSWAP. However, for all applications in this work, there is no need for a high-fidelity aSWAP gate. Instead, a significant portion of swap between the two components is all we need, which can be easily achieved by sweeping the flux bias of the coupler to go far beyond the relevant anticrossing point. For this purpose, a rough characterization such as \fig{circuit}(d) is already sufficient. This fact indicates robustness and efficiency of the aSWAP technique, which are welcome advantages for applications.

With the $Q_1-C_{12}-Q_2$ system characterized, we proceed to more detailed benchmark and control of the coupler using aSWAP operations. Figure\ \ref{spectrum}(b) shows an example histogram of measuring the coupler using the readout resonator of $Q_1$. The measure fidelity of 0.888 is very close to that of $Q_1$, indicating the high efficiency for coupler readout with the aSWAP operation. With this readout capability, we further characterize the coupler by performing measurements of Rabi oscillation (\fig{spectrum}(c)), relaxation process (\fig{spectrum}(d)), and Ramsey interference (\fig{spectrum}(e)). By analyzing these results, we can extract important parameters of the coupler such as decoherence times and driving amplitudes for gate operations.    

\begin{figure}[]
	\centering
	\includegraphics[width=\linewidth]{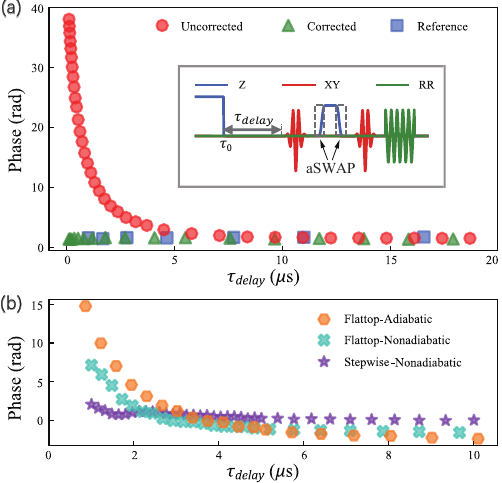}
	\caption{Calibration the flux distortion of a coupler. (a) Phase accumulation as a function of the delay time measured by a modified Ramsey interference. The uncorrected results exhibit a significant extra phase due to the flux distortion in the coupler up to several microseconds. The corrected results almost coincide with those obtained in a reference measurement in which no $Z$-pulse is applied and switched to zero at $\tau_0$ (thus no distortion at all), indicating high precise correction of the distortion. Inset: pulse sequence for the Ramsey interference. (b) A comparison of using adiabatic flattop (a duration of 50 ns for the rising and falling edges, respectively), nonadiabatic flattop (5 ns for the rising and falling edges, respectively), and stepwise (also nonadiabatic) pulses for the detection signal between the two $\pi/2$ pulses. The result shows that adiabaticity of the detection signal is critical for an accurate measurement of the accumulated phase. In these measurements, the frequency of the coupler is adjusted to its linear range during the flattop phase so that it is sensitive to the flux change.}
	\label{z-dist}
\end{figure}

Next, we demonstrate a precise, reliable, and straightforward approach for calibrating the coupler's flux distortion based on the aSWAP technique with fixed-frequency qubits. To accurately control the frequencies of couplers or qubits as desired, careful calibration and correction of flux distortion are required. For qubits, this task has been extensively investigated \cite{Rol:2020ewb,Sungprx,Zhiguang2019}. The key idea is to use a Ramsey sequence to detect the relative phase change in a qubit due to variations in its frequency. However, this strategy cannot be directly applied to a coupler since it does not have a dedicated readout resonator. In the following, we show that using our aSWAP technique, one can swap the states of a coupler and its nearby qubit and use the qubit's readout resonator to facilitate calibration and correction of flux distortion in the coupler. 

The inset of \fig{z-dist}(a) illustrates the principle of the calibration process. A rectangular flux pulse is abruptly switched to zero at time $\tau_0$. In the ideal case, the frequency of the coupler would also undergo an abrupt change accordingly. However, due to the distortion, the actual flux bias experienced by the coupler usually exhibits a long tail lasting up to a few microseconds after $\tau_0$. This distortion induces significant errors in subsequent operations involving the coupler. To characterize this distortion, we employ a modified Ramsey experiment to measure the extra phase accumulation on the coupler caused by the distortion at different delays \cite{Rol:2020ewb,Sungprx,Zhiguang2019}. To this end, a $\pi/2$ pulse is applied to a nearby qubit at time $\tau_{\textit{delay}}$ to initialize the coupler-qubit system into the state $(|0\rangle+|1\rangle)\otimes |0\rangle/\sqrt{2}$. Next, an aSWAP operation, depicted in \fig{circuit}(c), is used to exchange the states of the qubit and the coupler. The coupler is then left evolving under a constant flux bias to accumulate a phase $\phi$, where the coupler's frequency is sensitive to flux change. After that, a second aSWAP (the reverse operation of the first aSWAP) is performed to transform the state $|0\rangle \otimes (|0\rangle+e^{i\phi}|1\rangle)/\sqrt{2}$ into $(|0\rangle+e^{i\phi}|1\rangle)/\sqrt{2} \otimes |0\rangle$. Finally, a $\pi/2$ pulse is applied to the qubit, followed by a phase tomography measurement to complete the Ramsey experiment. We note that for this modified Ramsey measurement, the duration between the two $\pi/2$ pulses is fixed \cite{Rol:2020ewb,Sungprx,Zhiguang2019}. 

The phase accumulation, $\phi$, deduced from the Ramsey measurement (red circles in the main panel of \fig{z-dist}(a)) includes two contributions: the phase associated with the flattop bias, $\phi_0$, and the extra phase due to distortion, $\phi_{\textit{dist}}$. $\phi_0$ can be measured separately via a reference Ramsey experiment in which no pulse is applied to the coupler before $\tau_0$. $\phi_{\textit{dist}}$ can then be extracted as $\phi_{\textit{dist}}=\phi-\phi_0$. From \fig{z-dist}(a), it is clear that the distortion introduces a significant extra phase over an extended period. With this distortion information, we further apply a standard protocol to correct the distortion 
\cite{Zhiguang2019}. The negligible difference between the corrected case and the reference indicates that the distortion has been well corrected. 

We note that our calibration procedure differs from a standard Ramsey measurement \cite{Rol:2020ewb} in a critical way: It is performed on the qubit but is to measure the accumulated phase on the coupler. This distinction requires extra care when designing the experiment: A flattop flux pulse with aSWAP edges is applied to the coupler during the interval between the two $\pi/2$ rotations of the qubit. By doing this, the coupler is tuned to a region where its frequency is sensitive to flux change so that the accumulated phase is significant and easy to measure. Moreover, the rising and falling edges of this pulse serve as two aSWAP operations. Their adiabaticity is critical as we discuss below. 


Figure \ref{z-dist}(b) demonstrates the impact of adiabaticity by comparing the results of three measurements with rising and falling edges of different adiabaticity. For the stepwise-nonadiabatic case with sharp edges, the initial excited population of the qubit will be transferred to the upper branch of the spectrum near the anticrossing point (\fig{circuit}(c)) due to this nonadiabatic process. This means most of the excitation remains on the qubit instead of being swapped onto the coupler. As a result, the extra phase accumulation on the coupler is rather small and difficult to measure precisely. Even worse, a non-exponential tail appears in the measured data, possibly due to a Landau-Zener interference\cite{Campbell2020}. In contrast, when the edges are adiabatic, the excitation of the qubit will be swapped onto the coupler as desired, avoiding the above issues. For a flattop pulse with nonadiabatic edges, a phase accumulation of an intermediate magnitude results. Given the strong coupling between the qubit and the coupler (about 70 MHz in our case), the adiabaticity of an aSWAP operation can be easily satisfied when its duration exceeds a few tens of nanoseconds.

Compared to existing methods for calibrating flux distortion in couplers \cite{Sungprx, Collodo:2020btp, Zhangapl} , our method offers the following advantages. First, it does not require a dedicated or shared readout resonator for the coupler, making it readily applicable in most architectures of superconducting circuits that utilize tunable couplers. Second, single-qubit gates performed on the coupler are not necessary, eliminating the need for a precise characterization of the coupler itself. Third, aSWAP operations are typically much faster than operations that rely on $ZZ$ interaction between a qubit and a coupler used in other indirect methods \cite{Zhangapl}, so the influence of decoherence is less of a concern here. 

In the last part of this work, we demonstrate a new method for tuning effective qubit-resonator coupling using our aSWAP technique. In cQED architectures, the coupling between the qubit and resonator is described by the Jaynes-Cummings Hamiltonian \cite{Blais_MRP}, which can be approximated as $H=\left( \omega_\textit{r} +\chi \sigma_\textit{z} \right) \left( a^{\dagger} a+\frac{1}{2} \right)+\frac{\omega_{\textit{q}}}{2} \sigma_\textit{z}$ in the dispersive regime. Here $\omega_\textit{r}$ and $\omega_\textit{q}$ are the frequencies of the resonator and the qubit, respectively; $\sigma_\textit{z}$ is the Pauli operator acting on the qubit; $a$ and $a^{\dagger}$ are the annihilation and creation operators of the resonator's eigenmodes. The dispersive shift, $\chi=g^2/\Delta_{\textit{qr}}$, where $g$ is the coupling strength and $\Delta_{\textit{qr}}=\omega_\textit{r}-\omega_\textit{q}$ is the frequency detuning between the qubit and the resonator, quantifies the shift in the resonator frequency depends on the state of qubit.


The dispersive shift, $\chi$, is an important control parameter in many functional processes in cQED. For example, $\chi$ is adjusted by biasing the qubit frequency to optimize readout quality \cite{jeffrey2014fast, Arute_nature}. While a moderate $\chi$ compared to the readout resonator decay rate is usually chosen for strong measurement \cite{jeffrey2014fast,Arute_nature,Xu2020prl}, a smaller $\chi$ parameter is preferred for weak measurement, as this process intends to disturb the quantum system as little as possible \cite{ Vijay2012, dian2015}. In general, a large $\chi$ can result in extra decoherence in the qubit when the resonator's photon number is nonzero\cite{Clerk2010}. As a result, a tunable $\chi$ is highly desirable for the cQED platform. 

The coupling strength, $g$, is determined by geometric layout and is usually fixed in cQED circuits. Therefore, in architectures utilizing qubits with tunable frequencies, $\chi$ can be varied by adjusting the frequency detuning, $\Delta_{\textit{qr}}$. However, this method may be limited by factors such as frequency crowding and the presence of defects (e.g., two-level systems widely observed in solid-state devices). Moreover, tuning the frequencies of superconducting qubits is not feasible in certain cQED architectures \cite{Stehlik_PRL, Zhikun2024,chen2024leakage}. To address these issues, we propose and demonstrate a scheme that uses the aSWAP operations discussed above to adjust the dispersive shift between a qubit and its readout resonator in superconducting cQED circuits composed of qubits with fixed frequencies. This solution does not require additional components and can accurately tune the dispersive shift over a wide range.

\begin{figure}[t]
	\centering
	\includegraphics[width=\linewidth]{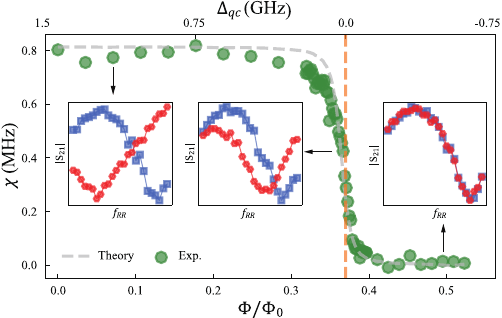}
	\caption{Dispersive shift of a readout resonator and its influence on the resonator's transmission coefficient.
                 The dispersive shift of the $Q_1$ readout resonator varies with the frequency of $C_{12}$(see \fig{circuit}). The gray dashed line is numerical simulation result and the green circles are experimental results. Insets: Changes in the readout resonator's transmission coefficient with the qubit at $|0\rangle$ (blue squares) and $|1\rangle$ (red hexagons) states as the dispersive shift varies as a function of the flux bias of the coupler.}
	\label{disper}
\end{figure}

In the tunable coupling structure depicted in \fig{circuit}(a), there is no direct coupling between the coupler and the readout resonator, and the indirect coupling is also negligible. Consequently, changing the state of the coupler has almost no effect on the resonator. On the one hand, this makes it difficult to directly read out the coupler using the resonator. On the other hand, it offers an alternative approach to controlling the dispersive shift $\chi$, as explained below. Due to the hybridization between the qubit and the coupler \cite{Blais2004}, the state $|\tilde{1}\rangle$ that we actually measure is a dressed state: $|\tilde{1}\rangle = \sin\theta |e0\rangle + \cos\theta |g1\rangle$, where $|e\rangle$ and $|g\rangle$ represent the first-excited and ground states of the coupler, respectively, and $|0\rangle$ and $|1\rangle$ are the bare states of the qubit. The angle $\theta$ is defined by $\tan\theta = g/(\Delta_{\textit{qc}}/2 - \sqrt{\Delta_{\textit{qc}}^2/4 + g^2})$, where $\Delta_{\textit{qc}}=\omega_c-\omega_q$ is the frequency detuning between the qubit and the coupler. For $\Delta_{\textit{qc}}>0$ and $\Delta_{\textit{qc}} \gg g$, $|\tilde{1}\rangle \approx |g1\rangle$, which means almost no hybridization occurs. Conversely, when $\Delta_{\textit{qc}} <0$ and $|\Delta_{\textit{qc}}|\gg g$, $|\tilde{1}\rangle \approx |e0\rangle$, indicating that the population is almost entirely swapped from the qubit to the coupler by an aSWAP operation. 

Such hybridization changes the dispersive shift from the original $\chi$ to the following effective value:
\begin{align}
    2\chi_{\textit{eff}} & =  \chi(\langle 0|\sigma_\textit{z}|0 \rangle - \langle \tilde{1}|\sigma_\textit{z}|\tilde{1}\rangle)\notag \\
    & = \chi\left(1-(\cos^2\theta-\sin^2\theta)\right) \\
    & = \chi \left(1-\left(\frac{-g^2 + (\Delta_{\textit{qc}}/2-\sqrt{\Delta_{\textit{qc}}^2/4 + g^2})^2}{g^2 + (\Delta_\textit{qc}/2-\sqrt{\Delta_\textit{qc}^2/4 + g^2})^2} \right) \right)\notag
\end{align}
$\chi_{\textit{eff}}\approx \chi$ for $\Delta_{\textit{qc}} >0$ and $\Delta_{\textit{qc}} \gg g$, and it decreases as $\Delta_{\textit{qc}}$ goes down. When $\Delta_{\textit{qc}} < 0$ and $|\Delta_{\textit{qc}}| \gg g$, $\chi_{\textit{eff}}$ approaches $0$, since in this regime, the population of the qubit is almost entirely swapped to the coupler via the aSWAP operation and the resonator cannot distinguish between the coupler's states. 

Figure\ \ref{disper} presents an experimental data set measuring the effective dispersive shift, $\chi_{\textit{eff}}$, of the $Q_1$-$C_{12}$ subsystem in \fig{circuit}(a) as a function of the flux bias applied to the coupler, which controls $\Delta_{\textit{qc}}$. As $\Delta_{\textit{qc}}$ transitions from positive to negative detuning, $\chi_{\textit{eff}}$ is continuously tuned from a finite value to nearly zero, effectively switching off the dispersive shift between $Q_1$ and its readout resonator. This substantial tunability renders the scheme potentially useful for the applications mentioned above. 


In summary, we have proposed and demonstrated a hardware-efficient method for robustly characterizing, calibrating, and controlling tunable couplers without the need for dedicated readout resonators. The core idea is to adiabatically swap the states of a coupler and its nearby qubit as desired, allowing the coupler's state to be readily measured using the qubit's readout resonator. The adiabaticity of the aSWAP operation can be easily achieved due to the strong coupling between the qubit and the coupler. To demonstrate applications of this aSWAP technique, we have used it to measure characteristic parameters of the coupler and calibrate its flux distortion. Furthermore, we show that it is possible to tune the dispersive shift between a qubit and its readout resonator in a wide range with this aSWAP operation, even though the qubit is frequency-fixed. Since tunable couplers without dedicated readout resonators are widely employed in mainstream superconducting quantum chips today, the aSWAP technique presented herein may become a useful addition to the toolbox for relevant research.

\noindent

\begin{acknowledgments}
This work was supported by the National Natural Science Foundation of China (No. 12074166), the Guangdong Provincial Key Laboratory (Grant No. 2019B121203002).
\end{acknowledgments}


	\vspace{10pt}
	\noindent


\begin{thebibliography}{25}%
\makeatletter
\providecommand \@ifxundefined [1]{%
 \@ifx{#1\undefined}
}%
\providecommand \@ifnum [1]{%
 \ifnum #1\expandafter \@firstoftwo
 \else \expandafter \@secondoftwo
 \fi
}%
\providecommand \@ifx [1]{%
 \ifx #1\expandafter \@firstoftwo
 \else \expandafter \@secondoftwo
 \fi
}%
\providecommand \natexlab [1]{#1}%
\providecommand \enquote  [1]{``#1''}%
\providecommand \bibnamefont  [1]{#1}%
\providecommand \bibfnamefont [1]{#1}%
\providecommand \citenamefont [1]{#1}%
\providecommand \href@noop [0]{\@secondoftwo}%
\providecommand \href [0]{\begingroup \@sanitize@url \@href}%
\providecommand \@href[1]{\@@startlink{#1}\@@href}%
\providecommand \@@href[1]{\endgroup#1\@@endlink}%
\providecommand \@sanitize@url [0]{\catcode `\\12\catcode `\$12\catcode `\&12\catcode `\#12\catcode `\^12\catcode `\_12\catcode `\%12\relax}%
\providecommand \@@startlink[1]{}%
\providecommand \@@endlink[0]{}%
\providecommand \url  [0]{\begingroup\@sanitize@url \@url }%
\providecommand \@url [1]{\endgroup\@href {#1}{\urlprefix }}%
\providecommand \urlprefix  [0]{URL }%
\providecommand \Eprint [0]{\href }%
\providecommand \doibase [0]{https://doi.org/}%
\providecommand \selectlanguage [0]{\@gobble}%
\providecommand \bibinfo  [0]{\@secondoftwo}%
\providecommand \bibfield  [0]{\@secondoftwo}%
\providecommand \translation [1]{[#1]}%
\providecommand \BibitemOpen [0]{}%
\providecommand \bibitemStop [0]{}%
\providecommand \bibitemNoStop [0]{.\EOS\space}%
\providecommand \EOS [0]{\spacefactor3000\relax}%
\providecommand \BibitemShut  [1]{\csname bibitem#1\endcsname}%
\let\auto@bib@innerbib\@empty
\bibitem [{\citenamefont {Blais}\ \emph {et~al.}(2021)\citenamefont {Blais}, \citenamefont {Grimsmo}, \citenamefont {Girvin},\ and\ \citenamefont {Wallraff}}]{Blais_MRP}%
  \BibitemOpen
  \bibfield  {author} {\bibinfo {author} {\bibfnamefont {A.}~\bibnamefont {Blais}}, \bibinfo {author} {\bibfnamefont {A.~L.}\ \bibnamefont {Grimsmo}}, \bibinfo {author} {\bibfnamefont {S.~M.}\ \bibnamefont {Girvin}},\ and\ \bibinfo {author} {\bibfnamefont {A.}~\bibnamefont {Wallraff}},\ }\bibfield  {title} {\bibinfo {title} {Circuit quantum electrodynamics},\ }\href {https://doi.org/10.1103/RevModPhys.93.025005} {\bibfield  {journal} {\bibinfo  {journal} {Rev. Mod. Phys.}\ }\textbf {\bibinfo {volume} {93}},\ \bibinfo {pages} {025005} (\bibinfo {year} {2021})}\BibitemShut {NoStop}%
\bibitem [{\citenamefont {{Google Quantum AI}}\ and\ \citenamefont {{Collaborators}}(2024)}]{Acharya2024}%
  \BibitemOpen
  \bibfield  {author} {\bibinfo {author} {\bibnamefont {{Google Quantum AI}}}\ and\ \bibinfo {author} {\bibnamefont {{Collaborators}}},\ }\bibfield  {title} {\bibinfo {title} {Quantum error correction below the surface code threshold},\ }\bibfield  {journal} {\bibinfo  {journal} {Nature}\ }\href {https://doi.org/10.1038/s41586-024-08449-y} {10.1038/s41586-024-08449-y} (\bibinfo {year} {2024})\BibitemShut {NoStop}%
\bibitem [{\citenamefont {Kim}\ \emph {et~al.}(2023)\citenamefont {Kim}, \citenamefont {Eddins}, \citenamefont {Anand}, \citenamefont {Wei}, \citenamefont {van~den Berg}, \citenamefont {Rosenblatt}, \citenamefont {Nayfeh}, \citenamefont {Wu}, \citenamefont {Zaletel}, \citenamefont {Temme},\ and\ \citenamefont {Kandala}}]{Kim2023}%
  \BibitemOpen
  \bibfield  {author} {\bibinfo {author} {\bibfnamefont {Y.}~\bibnamefont {Kim}}, \bibinfo {author} {\bibfnamefont {A.}~\bibnamefont {Eddins}}, \bibinfo {author} {\bibfnamefont {S.}~\bibnamefont {Anand}}, \bibinfo {author} {\bibfnamefont {K.~X.}\ \bibnamefont {Wei}}, \bibinfo {author} {\bibfnamefont {E.}~\bibnamefont {van~den Berg}}, \bibinfo {author} {\bibfnamefont {S.}~\bibnamefont {Rosenblatt}}, \bibinfo {author} {\bibfnamefont {H.}~\bibnamefont {Nayfeh}}, \bibinfo {author} {\bibfnamefont {Y.}~\bibnamefont {Wu}}, \bibinfo {author} {\bibfnamefont {M.}~\bibnamefont {Zaletel}}, \bibinfo {author} {\bibfnamefont {K.}~\bibnamefont {Temme}},\ and\ \bibinfo {author} {\bibfnamefont {A.}~\bibnamefont {Kandala}},\ }\bibfield  {title} {\bibinfo {title} {Evidence for the utility of quantum computing before fault tolerance},\ }\href {https://doi.org/10.1038/s41586-023-06096-3} {\bibfield  {journal} {\bibinfo  {journal} {Nature}\ }\textbf {\bibinfo {volume} {618}},\ \bibinfo {pages} {500} (\bibinfo {year}
  {2023})}\BibitemShut {NoStop}%
\bibitem [{\citenamefont {Arute}\ \emph {et~al.}(2019)\citenamefont {Arute}, \citenamefont {Arya}, \citenamefont {Babbush} \emph {et~al.}}]{Arute_nature}%
  \BibitemOpen
  \bibfield  {author} {\bibinfo {author} {\bibfnamefont {F.}~\bibnamefont {Arute}}, \bibinfo {author} {\bibfnamefont {K.}~\bibnamefont {Arya}}, \bibinfo {author} {\bibfnamefont {R.}~\bibnamefont {Babbush}}, \emph {et~al.},\ }\bibfield  {title} {\bibinfo {title} {Quantum supremacy using a programmable superconducting processor},\ }\href {https://doi.org/10.1038/s41586-019-1666-5} {\bibfield  {journal} {\bibinfo  {journal} {Nature}\ }\textbf {\bibinfo {volume} {574}},\ \bibinfo {pages} {505} (\bibinfo {year} {2019})}\BibitemShut {NoStop}%
\bibitem [{\citenamefont {Wu}\ \emph {et~al.}(2021)\citenamefont {Wu}, \citenamefont {Bao}, \citenamefont {Cao} \emph {et~al.}}]{Yulinprl}%
  \BibitemOpen
  \bibfield  {author} {\bibinfo {author} {\bibfnamefont {Y.}~\bibnamefont {Wu}}, \bibinfo {author} {\bibfnamefont {W.-S.}\ \bibnamefont {Bao}}, \bibinfo {author} {\bibfnamefont {S.}~\bibnamefont {Cao}}, \emph {et~al.},\ }\bibfield  {title} {\bibinfo {title} {Strong quantum computational advantage using a superconducting quantum processor},\ }\href {https://doi.org/10.1103/PhysRevLett.127.180501} {\bibfield  {journal} {\bibinfo  {journal} {Phys. Rev. Lett.}\ }\textbf {\bibinfo {volume} {127}},\ \bibinfo {pages} {180501} (\bibinfo {year} {2021})}\BibitemShut {NoStop}%
\bibitem [{\citenamefont {Stehlik}\ \emph {et~al.}(2021)\citenamefont {Stehlik}, \citenamefont {Zajac}, \citenamefont {Underwood}, \citenamefont {Phung}, \citenamefont {Blair}, \citenamefont {Carnevale}, \citenamefont {Klaus}, \citenamefont {Keefe}, \citenamefont {Carniol}, \citenamefont {Kumph}, \citenamefont {Steffen},\ and\ \citenamefont {Dial}}]{Stehlik_PRL}%
  \BibitemOpen
  \bibfield  {author} {\bibinfo {author} {\bibfnamefont {J.}~\bibnamefont {Stehlik}}, \bibinfo {author} {\bibfnamefont {D.~M.}\ \bibnamefont {Zajac}}, \bibinfo {author} {\bibfnamefont {D.~L.}\ \bibnamefont {Underwood}}, \bibinfo {author} {\bibfnamefont {T.}~\bibnamefont {Phung}}, \bibinfo {author} {\bibfnamefont {J.}~\bibnamefont {Blair}}, \bibinfo {author} {\bibfnamefont {S.}~\bibnamefont {Carnevale}}, \bibinfo {author} {\bibfnamefont {D.}~\bibnamefont {Klaus}}, \bibinfo {author} {\bibfnamefont {G.~A.}\ \bibnamefont {Keefe}}, \bibinfo {author} {\bibfnamefont {A.}~\bibnamefont {Carniol}}, \bibinfo {author} {\bibfnamefont {M.}~\bibnamefont {Kumph}}, \bibinfo {author} {\bibfnamefont {M.}~\bibnamefont {Steffen}},\ and\ \bibinfo {author} {\bibfnamefont {O.~E.}\ \bibnamefont {Dial}},\ }\bibfield  {title} {\bibinfo {title} {Tunable coupling architecture for fixed-frequency transmon superconducting qubits},\ }\href {https://doi.org/10.1103/PhysRevLett.127.080505} {\bibfield  {journal} {\bibinfo  {journal} {Phys.
  Rev. Lett.}\ }\textbf {\bibinfo {volume} {127}},\ \bibinfo {pages} {080505} (\bibinfo {year} {2021})}\BibitemShut {NoStop}%
\bibitem [{\citenamefont {Sung}\ \emph {et~al.}(2021)\citenamefont {Sung}, \citenamefont {Ding}, \citenamefont {Braum{\"u}ller} \emph {et~al.}}]{Sungprx}%
  \BibitemOpen
  \bibfield  {author} {\bibinfo {author} {\bibfnamefont {Y.}~\bibnamefont {Sung}}, \bibinfo {author} {\bibfnamefont {L.}~\bibnamefont {Ding}}, \bibinfo {author} {\bibfnamefont {J.}~\bibnamefont {Braum{\"u}ller}}, \emph {et~al.},\ }\bibfield  {title} {\bibinfo {title} {Realization of high-fidelity {{CZ}} and {{Z Z}} -free {{iSWAP}} gates with a tunable coupler},\ }\href {https://doi.org/10.1103/PhysRevX.11.021058} {\bibfield  {journal} {\bibinfo  {journal} {Phys. Rev. X}\ }\textbf {\bibinfo {volume} {11}},\ \bibinfo {pages} {21058} (\bibinfo {year} {2021})}\BibitemShut {NoStop}%
\bibitem [{\citenamefont {Yan}\ \emph {et~al.}(2018)\citenamefont {Yan}, \citenamefont {Krantz}, \citenamefont {Sung}, \citenamefont {Kjaergaard}, \citenamefont {Campbell}, \citenamefont {Orlando}, \citenamefont {Gustavsson},\ and\ \citenamefont {Oliver}}]{Yan2018prappld}%
  \BibitemOpen
  \bibfield  {author} {\bibinfo {author} {\bibfnamefont {F.}~\bibnamefont {Yan}}, \bibinfo {author} {\bibfnamefont {P.}~\bibnamefont {Krantz}}, \bibinfo {author} {\bibfnamefont {Y.}~\bibnamefont {Sung}}, \bibinfo {author} {\bibfnamefont {M.}~\bibnamefont {Kjaergaard}}, \bibinfo {author} {\bibfnamefont {D.~L.}\ \bibnamefont {Campbell}}, \bibinfo {author} {\bibfnamefont {T.~P.}\ \bibnamefont {Orlando}}, \bibinfo {author} {\bibfnamefont {S.}~\bibnamefont {Gustavsson}},\ and\ \bibinfo {author} {\bibfnamefont {W.~D.}\ \bibnamefont {Oliver}},\ }\bibfield  {title} {\bibinfo {title} {Tunable coupling scheme for implementing high-fidelity two-qubit gates},\ }\href {https://doi.org/10.1103/PhysRevApplied.10.054062} {\bibfield  {journal} {\bibinfo  {journal} {Phys. Rev. Appl.}\ }\textbf {\bibinfo {volume} {10}},\ \bibinfo {pages} {054062} (\bibinfo {year} {2018})}\BibitemShut {NoStop}%
\bibitem [{\citenamefont {Foxen}\ \emph {et~al.}(2020)\citenamefont {Foxen}, \citenamefont {Neill}, \citenamefont {Dunsworth} \emph {et~al.}}]{Foxen_prl}%
  \BibitemOpen
  \bibfield  {author} {\bibinfo {author} {\bibfnamefont {B.}~\bibnamefont {Foxen}}, \bibinfo {author} {\bibfnamefont {C.}~\bibnamefont {Neill}}, \bibinfo {author} {\bibfnamefont {A.}~\bibnamefont {Dunsworth}}, \emph {et~al.} (\bibinfo {collaboration} {Google AI Quantum}),\ }\bibfield  {title} {\bibinfo {title} {Demonstrating a continuous set of two-qubit gates for near-term quantum algorithms},\ }\href {https://doi.org/10.1103/PhysRevLett.125.120504} {\bibfield  {journal} {\bibinfo  {journal} {Phys. Rev. Lett.}\ }\textbf {\bibinfo {volume} {125}},\ \bibinfo {pages} {120504} (\bibinfo {year} {2020})}\BibitemShut {NoStop}%
\bibitem [{\citenamefont {Zhang}\ \emph {et~al.}(2022)\citenamefont {Zhang}, \citenamefont {Jiang}, \citenamefont {Deng} \emph {et~al.}}]{Zhang2022}%
  \BibitemOpen
  \bibfield  {author} {\bibinfo {author} {\bibfnamefont {X.}~\bibnamefont {Zhang}}, \bibinfo {author} {\bibfnamefont {W.}~\bibnamefont {Jiang}}, \bibinfo {author} {\bibfnamefont {J.}~\bibnamefont {Deng}}, \emph {et~al.},\ }\bibfield  {title} {\bibinfo {title} {Digital quantum simulation of floquet symmetry-protected topological phases},\ }\href {https://doi.org/10.1038/s41586-022-04854-3} {\bibfield  {journal} {\bibinfo  {journal} {Nature}\ }\textbf {\bibinfo {volume} {607}},\ \bibinfo {pages} {468} (\bibinfo {year} {2022})}\BibitemShut {NoStop}%
\bibitem [{\citenamefont {Xu}\ \emph {et~al.}(2020)\citenamefont {Xu}, \citenamefont {Chu}, \citenamefont {Yuan}, \citenamefont {Qiu}, \citenamefont {Zhou}, \citenamefont {Zhang}, \citenamefont {Tan}, \citenamefont {Yu}, \citenamefont {Liu}, \citenamefont {Li}, \citenamefont {Yan},\ and\ \citenamefont {Yu}}]{Xu2020prl}%
  \BibitemOpen
  \bibfield  {author} {\bibinfo {author} {\bibfnamefont {Y.}~\bibnamefont {Xu}}, \bibinfo {author} {\bibfnamefont {J.}~\bibnamefont {Chu}}, \bibinfo {author} {\bibfnamefont {J.}~\bibnamefont {Yuan}}, \bibinfo {author} {\bibfnamefont {J.}~\bibnamefont {Qiu}}, \bibinfo {author} {\bibfnamefont {Y.}~\bibnamefont {Zhou}}, \bibinfo {author} {\bibfnamefont {L.}~\bibnamefont {Zhang}}, \bibinfo {author} {\bibfnamefont {X.}~\bibnamefont {Tan}}, \bibinfo {author} {\bibfnamefont {Y.}~\bibnamefont {Yu}}, \bibinfo {author} {\bibfnamefont {S.}~\bibnamefont {Liu}}, \bibinfo {author} {\bibfnamefont {J.}~\bibnamefont {Li}}, \bibinfo {author} {\bibfnamefont {F.}~\bibnamefont {Yan}},\ and\ \bibinfo {author} {\bibfnamefont {D.}~\bibnamefont {Yu}},\ }\bibfield  {title} {\bibinfo {title} {High-fidelity, high-scalability two-qubit gate scheme for superconducting qubits},\ }\href {https://doi.org/10.1103/PhysRevLett.125.240503} {\bibfield  {journal} {\bibinfo  {journal} {Phys. Rev. Lett.}\ }\textbf {\bibinfo {volume} {125}},\
  \bibinfo {pages} {240503} (\bibinfo {year} {2020})}\BibitemShut {NoStop}%
\bibitem [{\citenamefont {Collodo}\ \emph {et~al.}(2020{\natexlab{a}})\citenamefont {Collodo}, \citenamefont {Herrmann}, \citenamefont {Lacroix}, \citenamefont {Andersen}, \citenamefont {Remm}, \citenamefont {Lazar}, \citenamefont {Besse}, \citenamefont {Walter}, \citenamefont {Wallraff},\ and\ \citenamefont {Eichler}}]{Collodo_prl}%
  \BibitemOpen
  \bibfield  {author} {\bibinfo {author} {\bibfnamefont {M.~C.}\ \bibnamefont {Collodo}}, \bibinfo {author} {\bibfnamefont {J.}~\bibnamefont {Herrmann}}, \bibinfo {author} {\bibfnamefont {N.}~\bibnamefont {Lacroix}}, \bibinfo {author} {\bibfnamefont {C.~K.}\ \bibnamefont {Andersen}}, \bibinfo {author} {\bibfnamefont {A.}~\bibnamefont {Remm}}, \bibinfo {author} {\bibfnamefont {S.}~\bibnamefont {Lazar}}, \bibinfo {author} {\bibfnamefont {J.-C.}\ \bibnamefont {Besse}}, \bibinfo {author} {\bibfnamefont {T.}~\bibnamefont {Walter}}, \bibinfo {author} {\bibfnamefont {A.}~\bibnamefont {Wallraff}},\ and\ \bibinfo {author} {\bibfnamefont {C.}~\bibnamefont {Eichler}},\ }\bibfield  {title} {\bibinfo {title} {Implementation of conditional phase gates based on tunable $zz$ interactions},\ }\href {https://doi.org/10.1103/PhysRevLett.125.240502} {\bibfield  {journal} {\bibinfo  {journal} {Phys. Rev. Lett.}\ }\textbf {\bibinfo {volume} {125}},\ \bibinfo {pages} {240502} (\bibinfo {year} {2020}{\natexlab{a}})}\BibitemShut
  {NoStop}%
\bibitem [{\citenamefont {Zhang}\ \emph {et~al.}(2023)\citenamefont {Zhang}, \citenamefont {Wang}, \citenamefont {Guo}, \citenamefont {Yang}, \citenamefont {Yang}, \citenamefont {Duan}, \citenamefont {Jia}, \citenamefont {Kong},\ and\ \citenamefont {Guo}}]{Zhangapl}%
  \BibitemOpen
  \bibfield  {author} {\bibinfo {author} {\bibfnamefont {C.}~\bibnamefont {Zhang}}, \bibinfo {author} {\bibfnamefont {T.-L.}\ \bibnamefont {Wang}}, \bibinfo {author} {\bibfnamefont {L.-L.}\ \bibnamefont {Guo}}, \bibinfo {author} {\bibfnamefont {X.-Y.}\ \bibnamefont {Yang}}, \bibinfo {author} {\bibfnamefont {X.-X.}\ \bibnamefont {Yang}}, \bibinfo {author} {\bibfnamefont {P.}~\bibnamefont {Duan}}, \bibinfo {author} {\bibfnamefont {Z.-L.}\ \bibnamefont {Jia}}, \bibinfo {author} {\bibfnamefont {W.-C.}\ \bibnamefont {Kong}},\ and\ \bibinfo {author} {\bibfnamefont {G.-P.}\ \bibnamefont {Guo}},\ }\bibfield  {title} {\bibinfo {title} {Characterization of tunable coupler without a dedicated readout resonator in superconducting circuits},\ }\href {https://doi.org/10.1063/5.0135219} {\bibfield  {journal} {\bibinfo  {journal} {Applied Physics Letters}\ }\textbf {\bibinfo {volume} {122}},\ \bibinfo {pages} {024001} (\bibinfo {year} {2023})}\BibitemShut {NoStop}%
\bibitem [{\citenamefont {Chen}\ \emph {et~al.}(2024)\citenamefont {Chen}, \citenamefont {Fors}, \citenamefont {Yan} \emph {et~al.}}]{chen2024leakage}%
  \BibitemOpen
  \bibfield  {author} {\bibinfo {author} {\bibfnamefont {L.}~\bibnamefont {Chen}}, \bibinfo {author} {\bibfnamefont {S.~P.}\ \bibnamefont {Fors}}, \bibinfo {author} {\bibfnamefont {Z.}~\bibnamefont {Yan}}, \emph {et~al.},\ }\href@noop {} {\bibinfo {title} {Fast unconditional reset and leakage reduction in fixed-frequency transmon qubits}} (\bibinfo {year} {2024}),\ \Eprint {https://arxiv.org/abs/2409.16748} {arXiv:2409.16748 [quant-ph]} \BibitemShut {NoStop}%
\bibitem [{\citenamefont {Luo}\ \emph {et~al.}(2023)\citenamefont {Luo}, \citenamefont {Huang}, \citenamefont {Tao}, \citenamefont {Zhang}, \citenamefont {Zhou}, \citenamefont {Chu}, \citenamefont {Liu}, \citenamefont {Wang}, \citenamefont {Cui}, \citenamefont {Liu}, \citenamefont {Yan}, \citenamefont {Yung}, \citenamefont {Chen}, \citenamefont {Yan},\ and\ \citenamefont {Yu}}]{luokai2023}%
  \BibitemOpen
  \bibfield  {author} {\bibinfo {author} {\bibfnamefont {K.}~\bibnamefont {Luo}}, \bibinfo {author} {\bibfnamefont {W.}~\bibnamefont {Huang}}, \bibinfo {author} {\bibfnamefont {Z.}~\bibnamefont {Tao}}, \bibinfo {author} {\bibfnamefont {L.}~\bibnamefont {Zhang}}, \bibinfo {author} {\bibfnamefont {Y.}~\bibnamefont {Zhou}}, \bibinfo {author} {\bibfnamefont {J.}~\bibnamefont {Chu}}, \bibinfo {author} {\bibfnamefont {W.}~\bibnamefont {Liu}}, \bibinfo {author} {\bibfnamefont {B.}~\bibnamefont {Wang}}, \bibinfo {author} {\bibfnamefont {J.}~\bibnamefont {Cui}}, \bibinfo {author} {\bibfnamefont {S.}~\bibnamefont {Liu}}, \bibinfo {author} {\bibfnamefont {F.}~\bibnamefont {Yan}}, \bibinfo {author} {\bibfnamefont {M.-H.}\ \bibnamefont {Yung}}, \bibinfo {author} {\bibfnamefont {Y.}~\bibnamefont {Chen}}, \bibinfo {author} {\bibfnamefont {T.}~\bibnamefont {Yan}},\ and\ \bibinfo {author} {\bibfnamefont {D.}~\bibnamefont {Yu}},\ }\bibfield  {title} {\bibinfo {title} {Experimental realization of two qutrits gate with tunable
  coupling in superconducting circuits},\ }\href {https://doi.org/10.1103/PhysRevLett.130.030603} {\bibfield  {journal} {\bibinfo  {journal} {Phys. Rev. Lett.}\ }\textbf {\bibinfo {volume} {130}},\ \bibinfo {pages} {030603} (\bibinfo {year} {2023})}\BibitemShut {NoStop}%
\bibitem [{\citenamefont {Rol}\ \emph {et~al.}(2020)\citenamefont {Rol}, \citenamefont {Ciorciaro}, \citenamefont {Malinowski}, \citenamefont {Tarasinski}, \citenamefont {Sagastizabal}, \citenamefont {Bultink}, \citenamefont {Salathe}, \citenamefont {Haandbaek}, \citenamefont {Sedivy},\ and\ \citenamefont {DiCarlo}}]{Rol:2020ewb}%
  \BibitemOpen
  \bibfield  {author} {\bibinfo {author} {\bibfnamefont {M.~A.}\ \bibnamefont {Rol}}, \bibinfo {author} {\bibfnamefont {L.}~\bibnamefont {Ciorciaro}}, \bibinfo {author} {\bibfnamefont {F.~K.}\ \bibnamefont {Malinowski}}, \bibinfo {author} {\bibfnamefont {B.~M.}\ \bibnamefont {Tarasinski}}, \bibinfo {author} {\bibfnamefont {R.~E.}\ \bibnamefont {Sagastizabal}}, \bibinfo {author} {\bibfnamefont {C.~C.}\ \bibnamefont {Bultink}}, \bibinfo {author} {\bibfnamefont {Y.}~\bibnamefont {Salathe}}, \bibinfo {author} {\bibfnamefont {N.}~\bibnamefont {Haandbaek}}, \bibinfo {author} {\bibfnamefont {J.}~\bibnamefont {Sedivy}},\ and\ \bibinfo {author} {\bibfnamefont {L.}~\bibnamefont {DiCarlo}},\ }\bibfield  {title} {\bibinfo {title} {Time-domain characterization and correction of on-chip distortion of control pulses in a quantum processor},\ }\href {https://doi.org/10.1063/1.5133894} {\bibfield  {journal} {\bibinfo  {journal} {Appl. Phys. Lett.}\ }\textbf {\bibinfo {volume} {116}},\ \bibinfo {pages} {54001} (\bibinfo {year}
  {2020})}\BibitemShut {NoStop}%
\bibitem [{\citenamefont {Yan}\ \emph {et~al.}(2019)\citenamefont {Yan}, \citenamefont {Zhang}, \citenamefont {Gong} \emph {et~al.}}]{Zhiguang2019}%
  \BibitemOpen
  \bibfield  {author} {\bibinfo {author} {\bibfnamefont {Z.}~\bibnamefont {Yan}}, \bibinfo {author} {\bibfnamefont {Y.-R.}\ \bibnamefont {Zhang}}, \bibinfo {author} {\bibfnamefont {M.}~\bibnamefont {Gong}}, \emph {et~al.},\ }\bibfield  {title} {\bibinfo {title} {Strongly correlated quantum walks with a 12-qubit superconducting processor},\ }\href {https://doi.org/10.1126/science.aaw1611} {\bibfield  {journal} {\bibinfo  {journal} {Science}\ }\textbf {\bibinfo {volume} {364}},\ \bibinfo {pages} {753} (\bibinfo {year} {2019})}\BibitemShut {NoStop}%
\bibitem [{\citenamefont {Campbell}\ \emph {et~al.}(2020)\citenamefont {Campbell}, \citenamefont {Shim}, \citenamefont {Kannan}, \citenamefont {Winik}, \citenamefont {Kim}, \citenamefont {Melville}, \citenamefont {Niedzielski}, \citenamefont {Yoder}, \citenamefont {Tahan}, \citenamefont {Gustavsson},\ and\ \citenamefont {Oliver}}]{Campbell2020}%
  \BibitemOpen
  \bibfield  {author} {\bibinfo {author} {\bibfnamefont {D.~L.}\ \bibnamefont {Campbell}}, \bibinfo {author} {\bibfnamefont {Y.-P.}\ \bibnamefont {Shim}}, \bibinfo {author} {\bibfnamefont {B.}~\bibnamefont {Kannan}}, \bibinfo {author} {\bibfnamefont {R.}~\bibnamefont {Winik}}, \bibinfo {author} {\bibfnamefont {D.~K.}\ \bibnamefont {Kim}}, \bibinfo {author} {\bibfnamefont {A.}~\bibnamefont {Melville}}, \bibinfo {author} {\bibfnamefont {B.~M.}\ \bibnamefont {Niedzielski}}, \bibinfo {author} {\bibfnamefont {J.~L.}\ \bibnamefont {Yoder}}, \bibinfo {author} {\bibfnamefont {C.}~\bibnamefont {Tahan}}, \bibinfo {author} {\bibfnamefont {S.}~\bibnamefont {Gustavsson}},\ and\ \bibinfo {author} {\bibfnamefont {W.~D.}\ \bibnamefont {Oliver}},\ }\bibfield  {title} {\bibinfo {title} {Universal nonadiabatic control of small-gap superconducting qubits},\ }\href {https://doi.org/10.1103/PhysRevX.10.041051} {\bibfield  {journal} {\bibinfo  {journal} {Phys. Rev. X}\ }\textbf {\bibinfo {volume} {10}},\ \bibinfo {pages} {041051}
  (\bibinfo {year} {2020})}\BibitemShut {NoStop}%
\bibitem [{\citenamefont {Collodo}\ \emph {et~al.}(2020{\natexlab{b}})\citenamefont {Collodo}, \citenamefont {Herrmann}, \citenamefont {Lacroix}, \citenamefont {Andersen}, \citenamefont {Remm}, \citenamefont {Lazar}, \citenamefont {Besse}, \citenamefont {Walter}, \citenamefont {Wallraff},\ and\ \citenamefont {Eichler}}]{Collodo:2020btp}%
  \BibitemOpen
  \bibfield  {author} {\bibinfo {author} {\bibfnamefont {M.~C.}\ \bibnamefont {Collodo}}, \bibinfo {author} {\bibfnamefont {J.}~\bibnamefont {Herrmann}}, \bibinfo {author} {\bibfnamefont {N.}~\bibnamefont {Lacroix}}, \bibinfo {author} {\bibfnamefont {C.~K.}\ \bibnamefont {Andersen}}, \bibinfo {author} {\bibfnamefont {A.}~\bibnamefont {Remm}}, \bibinfo {author} {\bibfnamefont {S.}~\bibnamefont {Lazar}}, \bibinfo {author} {\bibfnamefont {J.-C.}\ \bibnamefont {Besse}}, \bibinfo {author} {\bibfnamefont {T.}~\bibnamefont {Walter}}, \bibinfo {author} {\bibfnamefont {A.}~\bibnamefont {Wallraff}},\ and\ \bibinfo {author} {\bibfnamefont {C.}~\bibnamefont {Eichler}},\ }\bibfield  {title} {\bibinfo {title} {Implementation of conditional phase gates based on tunable $zz$ interactions},\ }\href {https://doi.org/10.1103/PhysRevLett.125.240502} {\bibfield  {journal} {\bibinfo  {journal} {Phys. Rev. Lett.}\ }\textbf {\bibinfo {volume} {125}},\ \bibinfo {pages} {240502} (\bibinfo {year} {2020}{\natexlab{b}})}\BibitemShut
  {NoStop}%
\bibitem [{\citenamefont {Jeffrey}\ \emph {et~al.}(2014)\citenamefont {Jeffrey}, \citenamefont {Sank}, \citenamefont {Mutus} \emph {et~al.}}]{jeffrey2014fast}%
  \BibitemOpen
  \bibfield  {author} {\bibinfo {author} {\bibfnamefont {E.}~\bibnamefont {Jeffrey}}, \bibinfo {author} {\bibfnamefont {D.}~\bibnamefont {Sank}}, \bibinfo {author} {\bibfnamefont {J.~Y.}\ \bibnamefont {Mutus}}, \emph {et~al.},\ }\bibfield  {title} {\bibinfo {title} {Fast accurate state measurement with superconducting qubits},\ }\href {https://doi.org/10.1103/PhysRevLett.112.190504} {\bibfield  {journal} {\bibinfo  {journal} {Phys. Rev. Lett.}\ }\textbf {\bibinfo {volume} {112}},\ \bibinfo {pages} {190504} (\bibinfo {year} {2014})}\BibitemShut {NoStop}%
\bibitem [{\citenamefont {Vijay}\ \emph {et~al.}(2012)\citenamefont {Vijay}, \citenamefont {Macklin}, \citenamefont {Slichter}, \citenamefont {Weber}, \citenamefont {Murch}, \citenamefont {Naik}, \citenamefont {Korotkov},\ and\ \citenamefont {Siddiqi}}]{Vijay2012}%
  \BibitemOpen
  \bibfield  {author} {\bibinfo {author} {\bibfnamefont {R.}~\bibnamefont {Vijay}}, \bibinfo {author} {\bibfnamefont {C.}~\bibnamefont {Macklin}}, \bibinfo {author} {\bibfnamefont {D.~H.}\ \bibnamefont {Slichter}}, \bibinfo {author} {\bibfnamefont {S.~J.}\ \bibnamefont {Weber}}, \bibinfo {author} {\bibfnamefont {K.~W.}\ \bibnamefont {Murch}}, \bibinfo {author} {\bibfnamefont {R.}~\bibnamefont {Naik}}, \bibinfo {author} {\bibfnamefont {A.~N.}\ \bibnamefont {Korotkov}},\ and\ \bibinfo {author} {\bibfnamefont {I.}~\bibnamefont {Siddiqi}},\ }\bibfield  {title} {\bibinfo {title} {Stabilizing rabi oscillations in a superconducting qubit using quantum feedback},\ }\href {https://doi.org/10.1038/nature11505} {\bibfield  {journal} {\bibinfo  {journal} {Nature}\ }\textbf {\bibinfo {volume} {490}},\ \bibinfo {pages} {77} (\bibinfo {year} {2012})}\BibitemShut {NoStop}%
\bibitem [{\citenamefont {Tan}\ \emph {et~al.}(2015)\citenamefont {Tan}, \citenamefont {Weber}, \citenamefont {Siddiqi}, \citenamefont {M\o{}lmer},\ and\ \citenamefont {Murch}}]{dian2015}%
  \BibitemOpen
  \bibfield  {author} {\bibinfo {author} {\bibfnamefont {D.}~\bibnamefont {Tan}}, \bibinfo {author} {\bibfnamefont {S.~J.}\ \bibnamefont {Weber}}, \bibinfo {author} {\bibfnamefont {I.}~\bibnamefont {Siddiqi}}, \bibinfo {author} {\bibfnamefont {K.}~\bibnamefont {M\o{}lmer}},\ and\ \bibinfo {author} {\bibfnamefont {K.~W.}\ \bibnamefont {Murch}},\ }\bibfield  {title} {\bibinfo {title} {Prediction and retrodiction for a continuously monitored superconducting qubit},\ }\href {https://doi.org/10.1103/PhysRevLett.114.090403} {\bibfield  {journal} {\bibinfo  {journal} {Phys. Rev. Lett.}\ }\textbf {\bibinfo {volume} {114}},\ \bibinfo {pages} {090403} (\bibinfo {year} {2015})}\BibitemShut {NoStop}%
\bibitem [{\citenamefont {Clerk}\ \emph {et~al.}(2010)\citenamefont {Clerk}, \citenamefont {Devoret}, \citenamefont {Girvin}, \citenamefont {Marquardt},\ and\ \citenamefont {Schoelkopf}}]{Clerk2010}%
  \BibitemOpen
  \bibfield  {author} {\bibinfo {author} {\bibfnamefont {A.~A.}\ \bibnamefont {Clerk}}, \bibinfo {author} {\bibfnamefont {M.~H.}\ \bibnamefont {Devoret}}, \bibinfo {author} {\bibfnamefont {S.~M.}\ \bibnamefont {Girvin}}, \bibinfo {author} {\bibfnamefont {F.}~\bibnamefont {Marquardt}},\ and\ \bibinfo {author} {\bibfnamefont {R.~J.}\ \bibnamefont {Schoelkopf}},\ }\bibfield  {title} {\bibinfo {title} {Introduction to quantum noise, measurement, and amplification},\ }\href {https://doi.org/10.1103/RevModPhys.82.1155} {\bibfield  {journal} {\bibinfo  {journal} {Rev. Mod. Phys.}\ }\textbf {\bibinfo {volume} {82}},\ \bibinfo {pages} {1155} (\bibinfo {year} {2010})}\BibitemShut {NoStop}%
\bibitem [{\citenamefont {Han}\ \emph {et~al.}(2024)\citenamefont {Han}, \citenamefont {Lyu}, \citenamefont {Zhou} \emph {et~al.}}]{Zhikun2024}%
  \BibitemOpen
  \bibfield  {author} {\bibinfo {author} {\bibfnamefont {Z.}~\bibnamefont {Han}}, \bibinfo {author} {\bibfnamefont {C.}~\bibnamefont {Lyu}}, \bibinfo {author} {\bibfnamefont {Y.}~\bibnamefont {Zhou}}, \emph {et~al.},\ }\bibfield  {title} {\bibinfo {title} {Multilevel variational spectroscopy using a programmable quantum simulator},\ }\href {https://doi.org/10.1103/PhysRevResearch.6.013015} {\bibfield  {journal} {\bibinfo  {journal} {Phys. Rev. Res.}\ }\textbf {\bibinfo {volume} {6}},\ \bibinfo {pages} {013015} (\bibinfo {year} {2024})}\BibitemShut {NoStop}%
\bibitem [{\citenamefont {Blais}\ \emph {et~al.}(2004)\citenamefont {Blais}, \citenamefont {Huang}, \citenamefont {Wallraff}, \citenamefont {Girvin},\ and\ \citenamefont {Schoelkopf}}]{Blais2004}%
  \BibitemOpen
  \bibfield  {author} {\bibinfo {author} {\bibfnamefont {A.}~\bibnamefont {Blais}}, \bibinfo {author} {\bibfnamefont {R.-S.}\ \bibnamefont {Huang}}, \bibinfo {author} {\bibfnamefont {A.}~\bibnamefont {Wallraff}}, \bibinfo {author} {\bibfnamefont {S.~M.}\ \bibnamefont {Girvin}},\ and\ \bibinfo {author} {\bibfnamefont {R.~J.}\ \bibnamefont {Schoelkopf}},\ }\bibfield  {title} {\bibinfo {title} {Cavity quantum electrodynamics for superconducting electrical circuits: An architecture for quantum computation},\ }\href {https://doi.org/10.1103/PhysRevA.69.062320} {\bibfield  {journal} {\bibinfo  {journal} {Phys. Rev. A}\ }\textbf {\bibinfo {volume} {69}},\ \bibinfo {pages} {062320} (\bibinfo {year} {2004})}\BibitemShut {NoStop}%
\end{thebibliography}%

%

\end{document}